\documentstyle[12pt]{article}
\makeatletter
\newcommand{\bd}{
\begin{document}}
\newcommand{\ed}{\end{document}}
\newcommand{\bc}{\begin{center}}
\newcommand{\ec}{\end{center}}
\newcommand{\be}{\begin{eqnarray}}
\newcommand{\ee}{\end{eqnarray}}
\newcommand{\nn}{\nonumber}
\newcommand{\eqn}{\global\def\theequation}
\newcommand{\mt}{m_t}
\newcommand{\mb}{m_b}
\newcommand{\ml}{m_l}
\newcommand{\me}{m_e}
\newcommand{\mm}{m_{\mu}}
\newcommand{\mte}{m_{\tau}}
\newcommand{\mgg}{M_{\gamma\gamma}}
%----------------------------------------------------------------------%
\bd
\begin{titlepage}
\begin{flushright}
 IS-J 4917\\ January~1993
\end{flushright}
%----------------------------------------------------------------------%
 \null
 \vskip 0.5in
\begin{center}
 \vspace{.15in}
{\Large {\bf   Right-handed Neutral Currents, Families }}
\\
\vspace{.10in}
{\Large {\bf and the LEP $l^+l^-\gamma\gamma$ Events}}
  \par
 \vskip 1.5em
 {\large
  \begin{tabular}[t]{c}
     C. Q. Geng,  K. Whisnant {\em and} B. -L. Young
\\
   \em  Department of Physics and Astronomy and Ames  Laboratory  \\
   \em  Iowa State University\\
   \em  Ames, IA 50011
  \end{tabular}}
 \par \vskip 5.0em
 {\large\bf Abstract}
\end{center}
\setlength{\baselineskip}{5ex}

We explore the possibility of having new physics which could
account for the $l^+l^-\gamma\gamma$ events with
$M_{\gamma\gamma}\simeq 60\ $GeV recently reported by LEP.
We consider models which contain an extra neutral gauge boson
($Z'$) that couples only to right-handed fermions.
The models naturally require at least three quark and lepton families from
anomaly cancellation. We find simple realizations of such models
that have $\ell^+\ell^-\gamma\gamma$ events at a rate similar to that
observed at LEP.

\vspace{1.0in}
\medskip
\vfill
\mbox{PACS\#: 12.15.Cc, 13.15.Jr.}

\end{titlepage}

\setlength{\baselineskip}{5ex}
\pagestyle{plain}
\pagenumbering{arabic}
\setcounter{page}{2}

Although the standard model of $SU(3)_C\times SU(2)_L\times U(1)_Y$
has been remarkably successful in explaining the existing experimental data,
it is generally expected that there is new physics beyond it
because of a large number of unanswered questions,
such as the origin of the intriguing family and mass patterns of the
quarks and leptons. Any experimental clue which could shed light on the
new physics would be extremely useful.

Recently, the L3 collaboration at LEP has reported [1] one
$e^+e^-\gamma\gamma$ and three $\mu^+\mu^-\gamma\gamma$ events with a
two photon invariant mass ($M_{\gamma\gamma}$) of
about $60\ $GeV, in a sample of $950,000$ Z's produced at
center-of-mass energies ranging from $88.2$ to $93.8\ $GeV.
These events cannot be explained by the standard model.
The DELPHI collaboration at LEP also has two similar events (one
$e^+e^-\gamma\gamma$ and one $\mu^+\mu^-\gamma\gamma$) [2].
However, some $l^+l^-\gamma\gamma$ events spreading over a
large range of $\gamma\gamma$ invariant masses were also observed [3].
Furthermore, all of the searches at LEP saw no $\nu\bar{\nu}\gamma\gamma$
event with $M_{\gamma\gamma}>10\ $GeV [1-3]. While it was estimated that two
events with a large lepton-photon angle and $M_{\gamma\gamma}>40$~GeV would be
generated by QED {\em bremsstrahlung} [3], the probability of having four such
events in a narrow mass bin was $O(10^{-3})$ [1]. It is clear
that more data are necessary to ascertain  the origin of these events.

In this letter, we will assume that at least some of the events with
$\mgg\simeq 60\ $GeV have a non-QED origin. It should be interesting to
explore the possible new physics which could account for the events; in
particular, we will examine scenarios which address the quark and lepton
family problem. Recently, Garisto and Ng [4] have discussed the
logical possibilities which could explain the events and pointed out that the
new physics which they may imply is extremely limited. The best possibility
is to have the $s$-channel process
\be
e^+e^-\:\to\: Z\  ({\tt on-shell})\
\to\: Z^{\prime*}\: (l^+l^-)\: X\:(\gamma\gamma)
\ee
where $Z'$ and $X$ are new spin 1 and 0 bosons, respectively, and the
asterisk indicates that the $Z^\prime$ is off-shell.
For the new particles, there are three main constraints:

(i) The coupling between $Z'$ and neutrinos should be suppressed
because of the non-observation of $\nu\bar{\nu}\gamma\gamma$ events.

(ii) The spin 0 boson $X$ with $M_X\simeq 60\ $GeV needs to decay
 mainly into $\gamma\gamma$.

(iii) The branching ratio of $Z\to l^+\l^-\gamma\gamma$ for each charged
lepton mode must be large enough to explain the measured rate.

The constraint (i) indicates that the $Z'$ cannot have a standard
type gauge boson coupling with leptons since the $Z$ branching fractions to
leptons obey $B(Z\to\nu\bar{\nu})/B(Z\to e^+e^-)\simeq 6$ (see [5]).
The most attractive way to satisfy (i) is to forbid the $Z'\nu\bar{\nu}$
coupling natually by a symmetry. Such a symmetry can be realized by adding
a gauged $U(1)'$ which has non-trivial charges only for the right-handed
fermions; then no light neutrino species will couple to the $U(1)'$ gauge
boson $Z'$.

There are two scenarios to accommodate (ii). One is to have
multiple Higgs doublets with the $X$ boson coming from a doublet which
decouples from the ordinary quarks and leptons but which still has couplings
to $W$, $Z$ and Higgs bosons [5]. The other is to allow couplings between
the $X$ boson and heavy charged fermions, such as the top-quark, a
fourth-generation of quarks and leptons, or certain exotic fermions.
In these two scenarios the $X$ boson decays mainly into $\gamma\gamma$
through virtual $W^{\pm}$ and heavy fermion loops, respectively. If
$B(X\to\gamma\gamma)\sim O(1)$, the rate of $Z\to l^+l^-\gamma\gamma$
depends on the strengths of the $ZZ'X$ and $Zl^+l^-$ couplings, and the
$Z'$ mass, $M_{Z'}$, which are constrained by various existing experiments.
In principle a large rate of  $Z\to l^+l^-\gamma\gamma$ may be achieved by
having large couplings  and a sufficiently small $M_{Z'}$. But, as we will
show below, it is very hard to build such a model in a simple way since the
couplings and $M_{Z'}$ are always related to each other in gauge theories, and
there are already strict constraints the couplings of extra $Z$ bosons from
$e^+e^-$ colliders such as TRISTAN [6,7].

Within the constraints described above we would like to construct
two explicit models based on the gauge group
$SU(3)_C\times SU(2)_L\times U(1)_Y\times U(1)'$
and examine how realistic the models would
be and their phenomenological implications. Here the $U(1)'$
is a local gauge symmetry which only couples with the right-handed quarks
and leptons and thus only $SU(2)_L$-singlet fermions may carry
$U(1)'$ charges. For simplicity we consider only cases in which
the representations of the quarks and leptons
under the gauge group $SU(3)_C\times SU(2)_L\times U(1)_Y$
are the standard ones, and electric charges are determined by the
usual relation $Q=I_{3L} + Y/2$.
There are four triangle anomaly-free conditions related
to the fermion charges of the $U(1)'$ from
$[SU(3)_C]^2\,U(1)'\,,\  [U(1)_Y]^2\,U(1)'\,,\
U(1)_Y\,[U(1)']^2$ and $[U(1)']^3$.
To ensure general covariance of the theory, the mixed chiral
gauge-gravitational anomaly of $U(1)'$ must be canceled.
This leads to one more condition that the sum of the $U(1)'$
charges vanishes, i.e., $TrQ'=0$. In addition to these theoretical
conditions, the models must satisfy the usual
mass requirements such as (a) the quark and charged-lepton
mass matrix determinants must be non-zero
%\footnote{We shall not discuss the case with $m_u=0$.}
and (b) there are only three light neutrino species and the mass
for any extra neutrino must be larger than $45\ $GeV [8].

We now present our models.
By allowing non-zero $U(1)'$ charges only for the right-handed quarks and
leptons, it is straightforward to show that at least three generations of
quarks and leptons are needed in order to satisfy the anomaly-free
conditions. Two minimal models (I) and (II) with three additional leptons,
$E_i\ (i=1,2,3)$ can be identified. In model (I), these three leptons are
simply three right-handed neutrinos. In model (II), $E_1$, $E_2$, and $E_3$
are exotic fermions, with the right-handed representations $(1,1,0)$,
$(1,1,2)$ and $(1,1,-2)$ under the standard group
$SU(3)_C\times SU(2)_L\times U(1)_Y$, and hence have electric charges 0, -1,
and 1, respectively. The fermion $U(1)'$ charge assignments are given by
\be
\matrix{
& u_R & c_R &  t_R & d_R & s_R & b_R & e_R & {\mu}_R & {\tau}_R
& E_{1R} & E_{2R} & E_{3R} \cr
& & &  & & & & & & && & \cr
Q'_I   & -1 & -1 & 2 & 1 & 1 & -2  & 1 &  1 &  -2 & -1 & -1 & 2\cr
& & &  & & & & & & && & \cr
Q'_{II}&  0 &  0 &  0 & 0 & 0 & 0  & -1 & -1 & 1 & -2 &  2 &  1\cr}\
\ee

Both models predict the existence of $\tau^+\tau^-\gamma\gamma$ events
which may be ignored presently because of harder experimental cuts. If
$\gamma\gamma +$ two-jet events are eventually observed above the QED
background, then model (I), in which the $Z^\prime$ also couples to quarks,
may be viable. Since no convincing evidence for such events currently exists,
we will concentrate on model (II) in the remainder of this letter. In this
model, to generate the fermion masses as well as to break the chiral
symmetries we introduce four Higgs doublets $\phi_i\ (i=1,2,3,4)$
with the following $U(1)'$ charges:
\be
\matrix{
&\phi_1 & \phi_2 & \phi_3 & \phi_4 \cr
& & & & \cr
Q^\prime_{II} & 0 & 1 & -1& Q^\prime \cr}\
\ee
As in the standard model, we assume that lepton number for each generation is
conserved. Thus the Yukawa couplings can be written as
\be
{\cal L}_Y =
\sum_{i,j}(h_{ij}^u\bar{q}_{L}^i\tilde{\phi}_1u_R^j+
h_{ij}^d\bar{q}_L^i\phi_1d_R^j)
\:+\:
h^e\bar{l}_{L}^e\phi_3e_R\:+\:
h^{\mu}\bar{l}_L^\mu\phi_3{\mu}_R\:+\:
h^{\tau}\bar{l}_L^\tau\phi_2{\tau}_R
\:+\: H.c.
\ee
where $i,j$ are generation indices and $\ell^\alpha_L$ are left-handed lepton
doublets ($\ell=e$, $\mu$ and $\tau$). The masses of the exotics $E_i$ can be
generated by introducing some Higgs singlets which transform as (1,1,0,$-$3)
and (1,1,0,4) under $SU(3)_c\times SU(2)_L\times U(1)_Y \times U(1)^\prime$.
Since these singlets will not affect our general discussion, we do not include
them in the rest of this letter. We note that since $\phi_4$ is decoupled from
the fermion sector, then its principal decay mode will be to two photons
through a $W$ loop.

Constraints on the $Z^\prime$ boson in model (II) come from $e^+e^-$
annihilation into charged lepton pairs. The vector and axial vector couplings
of the charged leptons to the $Z$ boson at 91.19~GeV have been precisely
measured for all three generations at LEP, and severely limit the size of
the $Z-Z^\prime$ mixing angle $\theta$. Taking the $Q^\prime_{II}$ charges
of Eq.~2, we find that the LEP limits give the approximate bound
\be
|\theta g_{Z^\prime}/g_Z| < 0.005,
\ee
where $g_{Z^\prime}$ is the $U(1)^\prime$
gauge coupling and $g_Z=e/(\sin\theta_W\cos\theta_W)$.

Further limits on the $Z^\prime$ can be obtained by analyzing $e^+e^-$ data
below the $Z$ resonance. Measurements of the total cross section and
forward-backward asymmetry in $e^+e^-\to \ell^+\ell^-$ ($\ell=\mu$ or $\tau$)
have been performed at TRISTAN for several values of $\sqrt{s}$ in the
range from 52~GeV to 64~GeV [6]. The TRISTAN data can be used to put a
lower limit on $M_{Z^\prime}$ for a given value of $g_{Z^\prime}$; the
relevant theoretical formalism is presented in Ref. [7]. Assuming negligible
$Z-Z^\prime$ mixing (as indicated by the LEP data), we find that in model (II)
for sufficiently large $M_{Z^\prime}$, this limit can be expressed as
\be
{g_{Z^\prime} M_Z\over M_{Z^\prime}g_Z} < {M_Z\over350{\rm~GeV}}=0.26,
\ee
at 95\%~CL.

One might ask how natural it is to have a small $Z-Z^\prime$ mixing angle.
This can be answered by examining the $Z-Z^\prime$ mass matrix in model (II)
\be
{\cal M}_Z^2 = {M_W^2\over\cos^2\theta_W}
\pmatrix{1
& 2(g_{Z^\prime}/g_Z)(v_2^2 - v_3^2 + Q^\prime v_4^2)/v^2 \cr
2(g_{Z^\prime}/g_Z)(v_2^2 - v_3^2 + Q^\prime v_4^2)/v^2
& 4(g_{Z^\prime}/g_Z)^2(v_2^2 + v_3^2 + Q^{\prime2}v_4^2)/v^2 \cr},
\ee
where $v_i$ are the vacuum expectation values (VEVs) of $\phi_i$ and
$v=\sqrt{v_1^2+v_2^2+v_3^2+v_4^2}$ is the standard model VEV. In Eq.~7 we
have not included possible contributions from singlet Higgs bosons in
$({\cal M}^2)_{22}$, since they do not affect our conclusions. If
$M_{Z^\prime}$ is sufficiently heavy then
\be
\theta {g_{Z^\prime}\over g_Z}\approx
2 {(v_2^2 - v_3^2 + Q^\prime v_4^2)\over v^2}
\left({g_{Z^\prime}\over M_{Z^\prime}} {M_Z\over g_Z}\right)^2.
\ee
Even when the TRISTAN limit of Eq.~6 is saturated, if there is a partial
cancellation of VEVs in the numerator (for example, to the 5\% level), then
$\theta g_{Z^\prime}/g_Z$ can be small enough to satisfy the LEP constraint.
We conclude that a small mixing angle is possible without a large amount of
fine tuning.

Given these limits on the $Z^\prime$, we can now calculate the rate for
$e^+e^-\to Z \to Z^{\prime*}(\ell^+\ell^-) X(\gamma\gamma)$ at the $Z$
resonance. If $X$ is the scalar component of $\phi_4$, the Higgs doublet that
does not couple to fermions, then $X$ will decay to two photons exclusively
through a $W$ loop. Also, since $Z^\prime$ does not couple to quarks or light
neutrinos, only the charged lepton pair mode is present. The rate relative to
the rate for a standard model Higgs boson of the same mass is given
approximately by
\be
R={(g_V^\prime)^2+(g_A^\prime)^2\over(g_V)^2+(g_A)^2}
\left({f_{ZZ^\prime X}\over f_{ZZH}}\right)^2
\left(M_Z\over M_{Z^\prime}\right)^4,
\ee
where $g_{V,A}^\prime$ and $g_{V,A}$ are the couplings of the charged leptons
to $Z^\prime$ and $Z$, respectively, $f_{ZZ^\prime X}$ is the $ZZ^\prime X$
coupling and $f_{ZZH}$ is the standard Higgs trilinear coupling to the $Z$
boson. The last factor in Eq.~9 comes from the $Z^\prime$ propagator compared
to the standard $Z$ propagator. Given the quantum numbers in Eqs.~2 and 3,
the ratio involving the lepton couplings is approximately
$8(g_{Z^\prime}/g_Z)^2$ and the ratio of Higgs couplings is
$(f_{ZZ^\prime X}/f_{ZZH})= 2 Q^\prime(v_4/v)(g_{Z^\prime}/g_Z)$;
then Eq.~9 becomes
\be
R=32\left({g_{Z^\prime}M_Z\over M_{Z^\prime}g_Z}\right)^4
\left(Q^\prime {v_4\over v}\right)^2.
\ee
The $Z^\prime$ mass must be at least as large as the $v_4$ contribution to
$({\cal M}^2)_{22}$ in Eq.~7, from which we deduce
\be
\left({g_{Z^\prime}M_Z\over M_{Z^\prime}g_Z}\right)
\left(Q^\prime {v_4\over v}\right) < {1\over2}.
\ee
Combining Eqs.~10 and 11 we find
\be
\left({g_{Z^\prime}M_Z\over M_{Z^\prime}g_Z}\right) > \sqrt{R\over8}.
\ee
The TRISTAN limit in Eq.~6 then implies that $R$ cannot be greater than
about 0.5 in model (II).

The standard model branching ratio for $Z\to e^+e^- H$ is
3.6$\times$10$^{-7}$ when $m_H =60$ GeV. The measured branching ratio for the
$\ell^+\ell^-\gamma\gamma$ events, averaged over the four LEP experiments [3],
is about 4$\times$10$^{-7}$ in each lepton channel, which would indicate an
$R$ value of order unity. Therefore we conclude that the model cannot account
for the $\ell^+\ell^-\gamma\gamma$ events at the current indicated rate if
all of the events are assumed to come from new physics. However, if the
measured rate were to drop by a factor of 2 or more, or some of the current
events are QED background, then the model would be viable.

The absence of $q\overline q\gamma\gamma$ and $\nu\overline\nu\gamma\gamma$
events provides a further constraint on the parameters, since in model (II)
there are also $e^+e^-\to Z \to Z^*(q\overline q) X(\gamma\gamma)$ and
$e^+e^-\to Z \to Z^*(\nu\overline\nu) X(\gamma\gamma)$ events. The branching
ratios for these processes are about 7$\times$10$^{-6}(v_4/v)^2$
and 2$\times$10$^{-6}(v_4/v)^2$, respectively. The nonobservation of
these modes imposes the approximate constraint $v_4/v < 1/3$. Combining this
limit with Eqs.~6 and 10, we see that in order to attain the maximum
$\ell^+\ell^-\gamma\gamma$ rate allowed by the TRISTAN data, $Q^\prime$ must
be 6 or more [9].

In a gauge theory the relevant couplings are determined solely by the gauge
coupling $g_{Z^\prime}$ and the $U(1)^\prime$ quantum numbers. Furthermore,
the same couplings which determine the rate for the virtual $Z^\prime$ going
into a lepton pair are always constrained by $e^+e^-$ collider data. Therefore
there should be a similar upper bound on the rate for
$\ell^+\ell^-\gamma\gamma$ events in gauge models which depend on the same
mechanism ($Z$ boson decaying to a virtual $Z^\prime$ boson and a Higgs boson
$X$ that does not couple to fermions).

While model (II) is the minimal extension in which the only standard
fermions that couple to the $Z^\prime$ boson are the charged leptons,
it is interesting to note that one can include an extra quark-lepton family
which has vanishing $U(1)^\prime$ charges (containing quarks $b^\prime$ and
$t^\prime$, charged lepton $\tau^\prime$ and left-handed neutrino
$\nu_{\tau^\prime}$) without upsetting the anomaly cancellation. The
exotic neutral fermion $E_1$ can be treated as a right-handed neutrino
associated with $\nu_{\tau^\prime}$, thereby making this fourth-generation
neutrino heavy and avoiding the LEP bound on the number of light neutrinos [8].
Then if the roles of $\tau$ and $\tau^\prime$ are reversed, the $Q^\prime_{II}$
charges of $e_R$, $\mu_R$, $\tau_R$ and $\tau^\prime_R$ become $-1$, $-1$,
$0$, and $1$, respectively. For this different set of $U(1)^\prime$ charges,
the TRISTAN data allows some solutions with $(g_{Z^\prime}m_Z/M_{Z^\prime}g_Z)
\approx 0.75$; in this modified model, the rate for $\ell^+\ell^-\gamma\gamma$
events via a virtual $Z^\prime$ can be much higher ($R\approx 4.5$) and
$Q^\prime$ (the $U(1)^\prime$ quantum number of $\phi_4$) can be as low as 2.
Also, there would be virtually no $\tau^+\tau^-\gamma\gamma$ events since
the $U(1)^\prime$ quantum number of the $\tau$ is zero (unlike model (II)).
Other extensions with more exotics can also be constructed; they typically
have maximum values of $R$ in the range from 0.1 to 10.

We conclude that if some or all of the $\ell^+\ell^-\gamma\gamma$ events are
not QED background and are due to new physics like that embodied in model (II),
then it is possible to have $Z\to Z^{\prime*}(\ell^+\ell^-) X(\gamma\gamma)$
events occurring at a rate similar to that observed at LEP. The models we
have studied naturally require three or four quark and lepton families.
In some, but not all, of the models $\tau^+\tau^-\gamma\gamma$ events should
eventually be seen at a similar rate. Barring an extreme
fine-tuning of  parameters, it is possible that indications of $Z-Z^\prime$
mixing would be seen at LEP~I as statistics are increased. Also, clear evidence
of $Z-Z^\prime$ interference should be detectable at LEP~II. Finally, there
are exotic fermions in the model which could perhaps be produced directly at
LEP~II.

\begin{center}
{\small {\bf ACKNOWLEDGEMENTS}}
\end{center}

This work was supported in part by the U. S. Department of Energy
under contract No. W-7405-Eng-82, Office of Energy Research
(KA-01-01), Division of High Energy and Nuclear Physics.

\vfill
\eject

\begin{center}
{\small {\bf REFERENCES}}
\end{center}

\noindent
\begin{enumerate}
\item L3 Collaboration, O. Adriani, et al.,
Phys. Lett. {\bf B295} (1992) 337.
\item J.~Marco, DELPHI report given at the DPF92, Fermilab, Batavia, Illinois,
(1992).
\item Cf. CERN Particle Physics Seminar on November 26, 1992 by
L3 Collaboration, followed by an open presentation by DELPHI, ALEPH and OPAL
Collaborations.
\item Robert Garisto and John N. Ng, TRIUMF Report No. TRI-PP-92-124.
\item V.~Barger N.~G.~Deshpande, J.~L.~Hewett and T.~G.~Rizzo, ANL Report
No. ANL-HEP-PR-92-102.
\item Talks by Y. Sugimoto, AMY collaboration, K. Hayashii, TOPAZ
collaboration, and J. Haba, VENUS Collaboration, at the 2nd KEK Topical
Conference on $e^+e^-$ Collision Physics, Tsukuba, Japan, Nov. 1991, KEK
Proceedings 92-9; talk by J. Shirai at the XXIst International Symposium on
Multiparticle Dynamics, Wuhan, China, Sept. 1991, KEK preprint 91-186.
\item K. Hagiwara, R. Najima, M. Sakuda, and N. Terunuma, Phys. Rev.
{\bf D41} (1990) 815.
\item Particle Data Group, Phys. Rev. {\bf D 45} (1992) S1.
\item This large value for $Q^\prime$ will give a large $Z^\prime Z^\prime XX$
coupling proportional to $(Q^\prime)^2$. Applying the condition
$f_{Z^\prime Z^\prime XX}/4\pi<1$, we get $g_{Z^\prime}/g_Z<0.9$, not an
unreasonable constraint.
\end{enumerate}
\ed